\documentstyle[preprint,aps,eqsecnum,tighten,pre,floats,epsfig,%
afterpage]{revtex}\long\def\yesPR#1{}

\def\notPR{}
\yesPR{\message{for PR}}\notPR

\newcommand{\lfr}[2]{{\textstyle{#1\over #2}}}	
\newcommand{\half}{{\textstyle{1\over2}}} 
\newcommand{\ie}{i.e.}\newcommand{\etal}{et al.}\newcommand{\etc}{etc.}
\newcommand{\via}{via}\newcommand{\eg}{e.g.}
\notPR\renewcommand{\ie}{{\it i.e.}}\renewcommand{\etal}{{\it et al.}}
\notPR\renewcommand{\etc}{{\it etc.}}\renewcommand{\via}{{\em via}}
\notPR\renewcommand{\eg}{{\em e.g.}}

\def\max{_{\rm max}}

\def\tfig#1{Fig.~\ref{#1}}
\notPR\def\ifigure#1#2#3#4{
\notPR\begin{figure}[t]\begin{center}{\epsfysize=#4truein \epsfbox{#3}}
\notPR\end{center}\smallskip
\notPR\caption{{\footnotesize #2}\label{#1}}\end{figure}}

\def\kbt{k_{\rm B}T}

\newcommand{\ex}[1]{{\rm e}^{#1}}                 
\newcommand{\fpd}[2]{{\delta #1\over\delta #2}}    
\newcommand{\ee}[1]{\cdot10^{#1}}\newcommand{\eem}[1]{\cdot10^{-#1}}
\newcommand{\inv}{^{\raise.15ex\hbox{${\scriptscriptstyle -}$}\kern-.05em 1}}
\newcommand{\pd}[2]{{\partial #1\over\partial #2}} 
\newcommand{\bnabla}{\mbox{\boldmath$\nabla$}}

\newcommand{\beq}{\begin{equation}}
\newcommand{\eeq}{\end{equation}}
\renewcommand{\b}[1]{{\bf #1}}
\newcommand{\lB}{\ell_B}
\newcommand{\ent}{_{{\rm ent}}}\newcommand{\es}{_{{\rm es}}}
\newcommand{\tot}{_{{\rm tot}}}
\newcommand{\self}{_{\rm self}}\newcommand{\gap}{_{\rm gap}}
\newcommand{\one}{_{\rm in}}

\newcommand{\np}{n_+}\newcommand{\nm}{n_-}\newcommand{\nf}{n_{{\rm
f}}}\newcommand{\npm}{n_\pm}
\newcommand{\vo}{v_0}
\newcommand{\sref}[1]{Sect.~\ref{#1}}
\newcommand{\eref}[1]{(\ref{#1})}
\newcommand{\nhat}{\hat n}
\newcommand{\fb}{\bar f}
\newcommand{\psb}{\bar\psi}
\newcommand{\sbar}{\bar\sigma}
\newcommand{\sbb}{\tilde\sigma}

\newcommand{\dd}{{\rm d}}

\newcommand{\smax}{\sigma_{\rm max}}
\newcommand{\smb}{\bar\sigma_-}\newcommand{\spb}{\bar\sigma_+}
\newcommand{\sob}{\bar\sigma\one}
\newcommand{\stb}{\bar\sigma_{\rm t}}\newcommand{\st}{\sigma_{\rm t}}
\newcommand{\spbb}{\tilde\sigma_+}

\newcommand{\fm}{f_{\rm m}}
\newcommand{\spav}{\sigma_{+,{\rm
av}}}\newcommand{\spavb}{\bar\spav}\newcommand{\stavb}{\bar\sigma_{\rm t,av}}
\def\offsetcurves{$ \fb/1000+10.9\spb-15.2$}
\def\offsetthreed{$- \fb(\spb,\stb)/1000+218.0+6.0\spb-11.4\stb$}
\def\offsetthick{$\fb-12750+5615\spb$}

\begin{document}
\draft

\title{Charge-Reversal Instability in Mixed Bilayer Vesicles}
\author{Yi Chen and Philip Nelson}
\address{Department of Physics and Astronomy, University of Pennsylvania,
Philadelphia PA 19104}
\date{16 February 2000}
\maketitle
\begin{abstract}
Bilayer vesicles form readily from mixtures of charged and neutral
surfactants. When such a mixed vesicle binds an oppositely-charged
object, its membrane partially demixes: the adhesion zone recruits
more charged surfactants from the rest of the membrane. Given an
unlimited supply of adhering objects one might expect the vesicle to
remain attractive until it was completely covered. Contrary to this
expectation, we show that a vesicle can instead exhibit {\it adhesion
saturation,} partitioning spontaneously into an attractive zone with
definite area fraction, and a repulsive zone. The latter zone rejects
additional incoming objects because counterions on the interior of the
vesicle migrate there, effectively reversing the membrane's charge.
The effect is strongest at high surface charge
densities, low ionic strength, and with thin, impermeable
membranes. Adhesion saturation in such a situation has recently been
observed experimentally [H. Aranda-Espinoza {\it et al.}, {\sl
Science}  {\bf285} 394--397 (1999)].

\end{abstract}

\notPR\renewcommand{\baselinestretch}{1.2}
\section{Introduction}\label{sintro}
The self-assembly of colloidal particles offers an attractive route to
the synthesis of highly ordered, nanostructured materials. Typically
these materials have been extremely soft, being stabilized by entropic
effects.

For example, classical colloidal crystals are three-dimensional arrays
of mutually repelling spheres \cite{hach73a,will76a}. Entropic effects maintain their
crystalline order in spite of a density well below that of
close-packing. As a result, these arrays are easily disrupted by small
mechanical shear, dilution, \etc{}  More recently, depletion forces
have been harnessed to assemble spheres into crystalline arrays on the
walls of their container \cite{dins97a}. Again the physical forces between
the spheres are repulsive, and again the resulting arrays are
extremely soft.

Attempts to create strong ordered materials from physically {\it
attracting} components have generally produced instead highly
disordered aggregates. Recently, however, Ramos \etal{} reported
the observation of robust two-dimensional
crystallites formed from negatively-charged latex spheres introduced
into a suspension of bilayer vesicles \cite{aran99a,ramo99a}. The membranes
forming the
vesicles consist of a mixture of positively-charged and neutral
surfactants. The immense electrostatic attraction between the negative
spheres and positive membranes led to the crystallites' great
strength; their ordered 2d character arose \via{} the intermediary role
of the vesicles as {\it templates} for the initial self-assembly of the
spheres.

In this paper we develop some of the physics of the crucial
intermediate step just mentioned, elaborating and extending the
discussion in \cite{aran99a}.  This stage begins
when the latex spheres are first introduced to the vesicle suspension,
and lasts for hours to days. Initially the spheres adsorb avidly onto
the vesicles, and indeed many vesicles become completely covered with
spheres. However, a significant subpopulation of vesicles content
themselves with only partial coverage: on these vesicles the adsorbed
spheres form a {\it self-limiting `raft'.} Once the raft forms, no
further spheres attach to the vesicle anywhere, though they are
present in excess. Instead, particles in suspension are
seen to approach, then wander away from, the vesicle.

The theory of colloidal surface interactions is vast (for
introductions see \cite{russelbook,israelachvilibook,safranbook}).
Our goal is to introduce a very simple mechanism for adhesion
saturation, summarized graphically in \tfig{fsetup} below, then
present some calculations to show how it works in the parameter regime
relevant to experiments. We will argue that our effect should be
qualitatively unchanged after many other surface-interaction effects
are included in the analysis, but much work remains to be done to show
this in detail. \sref{spp} sketches the physics of our
mechanism. \sref{slt} begins the analysis using linearized
Poisson--Boltzmann theory, considering in turn a series of more
complicated situations. The linearized theory is familiar and helps to
connect the analysis to the physical picture, but it proves to be
inadequate for the interesting range of parameter values. Thus in
\sref{snpbt} we upgrade to the full nonlinear theory, which proves to be
quite easy in this context. Finally we consider the effects of ion
correlations, neglected in Poisson--Boltzmann theory, in \sref{seic}.
A glossary of symbols appears in the Appendix.

\section{Physical Picture}\label{spp}
We first briefly review the physical picture developed in
\cite{aran99a} and summarized in \tfig{fsetup}.

Consider first two dielectric surfaces bearing fixed charge densities
$\sigma_\pm$ of the same magnitude but opposite sign in an electrolyte
solution. When they are separated by several screening lengths they
feel little mutual attraction, since each maintains a neutralizing
cloud of counterions. As the surfaces approach closer, eventually
their screening clouds begin to interpenetrate. Then negative
counterions from the positive surface, and positive counterions from
the negative surface, can escape to infinity without violating overall
charge neutrality. The corresponding gain in entropy reduces the
system's free energy: counterion release drives the surfaces into
contact.

\def\capone{{\it (a)~}~Cartoon of the situation. A large vesicle of
mixed neutral and positively-charged surfactants attracts a limited
number of negatively-charged spheres, then saturates. The Debye
screening length, typically about 10~nm, is much smaller than the sizes of the
objects.\hfil\break
{\it (b)~}~Disposition of counterions when an approaching negative object
(shaded, above right) is still far from the
vesicle. The vesicle interior is at the bottom of the figure. The
zeros denote neutral surfactants, plus signs the charged
surfactants. Circled $\pm$ signs denote counterions in solution. The
solid vertical lines joining
charges are fictitious elastic tethers representing intuitively the
electric field lines; the requirement of charge neutrality translates
visually into the requirement that all charges be tied in this way.\hfil\break
{\it (c)~}~Redistribution of charges when the negative dielectric object
approaches the membrane, if we artificially forbid any electric field
inside the membrane. Four pairs of counterions have been released to infinity
(upper left). The interior monolayer, and its counterion
cloud, are unchanged from {\it(b)}. Zone ``n''  presents a net of one
positive charge to the vesicle exterior and so remains attractive to further
incoming negative objects.\hfil\break
{\it (d)~}~The resulting state after we relax the constraint of zero
electric field inside the membrane, allowing the ion migrations indicated
by the horizontal dashed arrows in {\it(c)}. One additional counterion pair has
been
released to infinity and the adhesion gap has narrowed. The net charge
of the bilayer plus interior counterions in zone ``n'' has reversed sign
relative to
{\it (c)}, and so this zone repels additional incoming negative objects.
[Adapted with permission from \cite{aran99a}. \copyright{} 1999
American Association for the Advancement of Science.]}

\notPR\ifigure{fsetup}{\protect\capone}{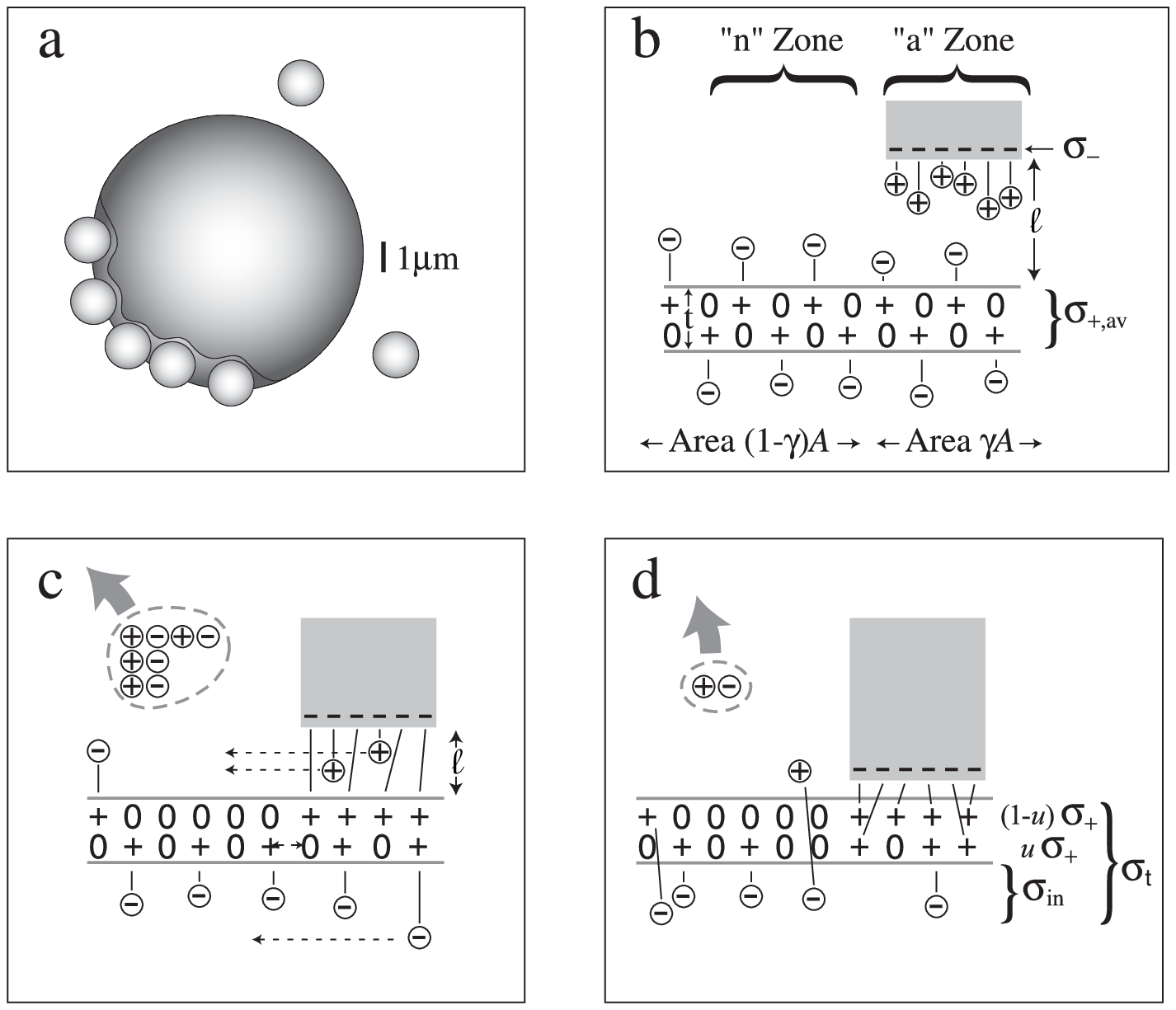}{4}\afterpage{\clearpage}

Next consider the case of two surfaces of opposite sign and {\it
unequal} magnitude; for instance, suppose that
$\sigma_+<|\sigma_-|$. In this case counterion release will be
incomplete; after exhausting all the negative counterions, some
positive ones will remain, trapped by the requirement of charge
neutrality. The osmotic pressure of the trapped ions will prevent
the surfaces from coming into perfect contact. If one surface has {\it
variable} charge density, say $\sigma_+$, then additional surface
charges will be pulled into the contact region in order to improve the
contact with the approaching negative surface \cite{nard98a}. A surface charge
density can for instance vary because the {\it composition} of the
surface is variable: for instance, the surface may be a mixture of
charged and neutral surfactants, as in the experiments of
\cite{aran99a,ramo99a}. In this case the recruitment of charge to the contact
region will deplete the other regions, in turn rendering them less
attractive to additional negative dielectric
objects. \tfig{fsetup}{\it c} depicts this situation: surfactant
rearrangement in the outer monolayer of the membrane has permitted the
release of two more ion pairs than would otherwise (panel {\it b}) have been
possible.

The rearrangement of membrane charges is limited: the relative
concentration of charged surfactants cannot exceed unity.  The maximum
charge density on the outer monolayer may still be less than that of
the approaching dielectric, and so the final contact may still be
imperfect, as shown in \tfig{fsetup}{\it c}. We will assume this to be
the case in the rest of this paper.

Nevertheless, a further reduction
in free energy density from  \tfig{fsetup}{\it c} is still possible,
once we remember that the inner membrane monolayer and its counterions
need not play a passive role. \tfig{fsetup}{\it d} shows how the
remaining trapped counterions in panel {\it c} can leave the gap, even
if the membrane is impermeable, by following the dashed horizontal
arrows in panel {\it c}. After this rearrangement some of the charge
on the negative dielectric is neutralized by surfactants on the inner
monolayer, whose own interior counterions migrate to the nonadhesion
region%
\notPR .\footnote{Even if the dielectric's charge exceeds twice the
\notPR monolayer charge density, as assumed in the text below,
\notPR additional $\pm$ ion pairs can be brought from the membrane interior,
\notPR with the positive ions remaining in the adhesion region to help
\notPR neutralize the dielectric and the negative ones migrating to the
\notPR nonadhesion region, driving its net charge still more negative.}
\yesPR{~\cite{fna}.}
Panel {\it c} also shows a rearrangement of the surfactants on the
{\it inner} monolayer, further depleting the charge of the noncontact
zone.

\tfig{fsetup}{\it d} raises an intriguing question: will the migration
of interior counterions ever overwhelm and effectively reverse the
charge of the membrane as seen from outside, as shown in the figure?
Of course, cartoons alone will not settle this question, but we can
argue physically that such an effect may well happen as follows. First
we note that the positive charge density of the noncontact zone ``n''
is already very small in \tfig{fsetup}{\it c}, since the
interior monolayer and its counterion cloud cancel, and as we will see
below the electrostatic interaction driving the depletion of charged
surfactants from the outer monolayer is very strong. Thus only a small
migration of interior counterions will suffice to get charge
reversal. Second, the entropic {\it cost} of creating a nonuniform charge
density in the interior counterion cloud is quadratic in the amount of
charge which migrates, since the uniform distribution is an
equilibrium state. But the free energy {\it gain} from this
redistribution is  linear in the amount of charge migration, being dominated by
the derivative $\dd/\dd\sigma_+$ of the attractive self-energy
(see formula \eref{edffgapLPB} below). Thus a finite amount of counterion
migration will occur, and this amount may well exceed the small net
charge on the nonadhesion region, effectively reversing it.

The rest of this paper is devoted to a quantitative justification of
the intuitive argument just given. Before passing on to the analysis,
we should remark on another feature of \tfig{fsetup}{\it d}. Charge
reversal requires that electric fields (represented schematically by
the vertical lines in the figure) penetrate the interior of the
membrane. Since the membrane interior is a low dielectric constant
medium, the energetic cost of these fields can be significant, another
term quadratic in the amount of charge migration from panel {\it c} to {\it
d}. If the membrane is sufficiently thick, this cost will reduce the
charge migration below the point of charge reversal, a point we will
need to examine quantitatively in \sref{ssftbm.npbt} below.

\section{Linearized Mean-Field Theory}\label{slt}
In this section we begin the mathematical implementation of the
ideas in \sref{spp}. We begin with the linearized (Debye-H\"uckel)
limit of low charge density, even though ultimately we will argue that
the experiments studied here require a full nonlinear treatment. We do
this partly because of the simplicity of the formul\ae, and partly to
make contact with earlier work.

To fix notation and  keep the article self-contained we begin by
rederiving some key results from \cite{pars72a,nard97a,nard98a}.
The Appendix summarizes our units and all symbols used throughout the
paper.

\subsection{Basic formul\ae}\label{ssbf.lt}
The electrostatic potential energy of a distribution of free charges of
density $\rho(\b r)$ is
\( \half\int\dd\b r\,\rho(\b r)\psi(\b r) \), where $\psi$ is the
electric potential \cite{landaumediabook}. The potential created by a
single point charge
\(q\) in an infinite, uniform, dielectric medium is \( \psi(\b
r)=q/4\pi\epsilon|\b r| \). In a more complicated situation, \(
\psi(\b r) \) is related to \( \rho(\b r') \) by some Green function
\( G(\b r,\b r') \) and obeys Poisson's equation,
$\nabla^2\psi=-\rho/\epsilon$.

We first imagine a uniform charge distribution of density \( \sigma \)
on the surface \( \{z=0\} \)%
\notPR .\footnote{The assumption of fixed charge
\notPR is appropriate for  surfaces with fully-ionized groups at the pH used,
\notPR such as those in the experiments of \cite{aran99a,ramo99a}. We also implicitly
\notPR assume that the surfactants used are insoluble in water, so that their
\notPR numbers in the membrane are fixed. This assumption may need
\notPR further scrutiny,  since in the experiments one surfactant species forms micelles.} 
\yesPR{~\cite{fnb}.}
The halfspace \( z<0 \) is filled with a
dielectric with no free charges, and so the electric field must
everywhere vanish here. The other halfspace is a univalent salt solution in
equilibrium with a reservoir at concentration $\nhat$. The
reservoir must remain neutral, but it can supply ion pairs at a cost
in free energy given by a chemical potential $\mu\kbt$.
The total free energy of the mobile ions near the surface is
then \( F=F\ent+F\es \), where the entropic and electrostatic energies
in mean-field approximation are \cite{safranbook}
\beq
F\ent=\kbt\int\dd\b r\,\left[
\np(\ln\np\vo - 1)+\nm(\ln\nm\vo - 1)-\mu(\np+\nm)+\xi(\np-\nm+\nf)
\right]
\label{edfFent}\eeq
\beq
F\es=\half\int\dd\b r\dd\b r'\,en\tot(\b r) G(\b r,\b r')en\tot(\b r')
\ .\label{edfFes}\eeq
In the above formul\ae{}, $n_\pm$ are the number densities of ions,
while \( n\tot(\b r)=\rho/e=\np-\nm+\nf \) is the total signed
density, including fixed surface charges with signed density \( \nf
\). We introduced a Lagrange 
multiplier $\xi$ to enforce overall neutrality. The symbol $v_0$ is a
microscopic volume factor which will drop out of all physical
results. We have fixed the arbitrary constant in $F\es$ by setting
the electrostatic energy to zero when
the mobile counterions form a sheet coinciding with the fixed surface
charge. Thus $F\es$ is the work needed to pull this sheet away from
the surface, and so is a positive quantity.

In equilibrium we have \( \fpd F{\npm(\b r)}=0 \). Away from the plane
this fixes
\beq \npm(\b r)\vo=\ex{\mu\mp(\psb(\b r)+\xi)}\ ,\quad z>0\ .\eeq
Here $\psb=e\psi/\kbt$ and we have fixed the additive constant in \(
\psi \) by choosing $\psi(\infty)=0$. Since \( \np=\nm=\nhat \) at
infinity, we get $\xi=0$ and \( \mu=\ln \nhat\vo \), or
\begin{equation} \npm(\b r)= \nhat\ex{\mp\psb(\b r)}
\ .\label{enpm}\label{edensity}\end{equation}
Substituting then gives the free energy
\beq F=\kbt\nhat\int\dd\b r\,\left[
\psb\sinh\psb-2\cosh\psb+\half(\nf/\nhat)\psb
+2
\right]
\ .\label{edfFfull}\eeq
The last term of \eref{edfFfull} is a constant which we have added by
hand to cancel a term proportional to the volume of the world.

Eqn.~\eref{edfFfull} simplifies if the dimensionless potential $\psb$ is
everywhere $\ll1$; in this case we simply get \( F=\kbt\int\dd\b
r\,\half \nf\psb \). Since the fixed charge $n_f$ is confined to a plane,
the free energy is a purely surface term once $\psb$ has been found.

To find \( \psb \), we note that it satisfies the Poisson equation, a
property of the Green function used to define it. Using the charge
density \( en_\pm(\b r) \) found above in \eref{enpm} gives  the
Poisson-Boltzmann
equation,
\beq\nabla^2\psb={2e^2\nhat\over\epsilon\kbt}\sinh\psb
\ .\label{ePBefull}\eeq
Linearizing then gives the familiar Debye-H\"uckel equation:
\( \nabla^2\psb=\kappa^2\psb \), where \( \kappa=\sqrt{2e^2\nhat/
\epsilon\kbt} \).

The objects we want to consider are much bigger than the screening
length $\lambda_D=1/\kappa$ (see \tfig{fsetup}{\it a}\/). Thus our geometry is
essentially planar, and we need the planar solutions \( \psb(z)=B
\ex{\pm\kappa z} \) to the Debye-Huckel equation.  The electric field
is then $\b E=-\bnabla\psi$, which indeed
decays exponentially on the length scale $\lambda_D$.

For a single wall we must choose the decaying solution to
\eref{ePBefull}.
We fix the constant $B$ by imposing Gauss's law at the surface: \(
\b E=-\pd\psi z\hat{\b z}={\sigma\over\epsilon}\hat{\b z} \). Then
\( B=\sigma e/\kappa\epsilon\kbt\), the solution is
\beq \psb(z)={\sigma e\over\kappa\epsilon\kbt}\ex{-\kappa z}
\ .\qquad\ {\rm (linearized\ approximation)}\label{edfsolDH}\eeq
and the free energy per unit area
of the isolated, charged surface is
\beq
f\self\equiv F/{\rm (area)}=\kbt\sigma B/2e={\sigma}^2/2\kappa\epsilon
\ .\qquad{\rm (linearized\ approximation)}
\label{edffselfDH}\eeq

Another well-known solution to \eref{ePBefull} arises in the opposite
case of very high charge density, where \(\psb\gg 1 \) at the surface. In this
case the
Poisson-Boltzmann equation has a solution of ``Gouy-Chapman'' form:
\(\psb(z)=\ln\left[{2\epsilon\kbt\over e^2\nhat}{1\over(z+\lambda_{\rm
GC})^2}\right] \). Here
the free parameter is the offset $\lambda_{\rm GC}$, chosen to enforce Gauss's
law:
$\lambda_{\rm GC}=2\epsilon\kbt/e\sigma$. More highly-charged surfaces thus
have
smaller $\lambda_{\rm GC}$ and so a
more nearly singular potential. The pathological behavior of $\psb$ at
large $z$ simply reflects the end of the regime \(\psb\gg1 \) at large
enough $z$. Note that the electric field $E_z=2\kbt/e(z+\lambda_{\rm GC})$ of
the
Gouy-Chapman solution is independent of the ambient salt concentration
$\nhat$, as it should be: the electric forces near a highly
charged surface depend only on the surface charge. The salt
concentration determines only the extent of the region in which the
strong-field approximation is valid.

\subsection{Two dielectrics}\label{sstd.lt}
We minimized the free energy of an isolated surface, obtaining
\eref{edffselfDH}. To extract any useful work from this stored free
energy, we would have to remove some constraint. One way to do this is
to bring in another semiinfinite, planar dielectric%
\notPR \footnote{Nothing
\notPR is really infinite. The phrase
\notPR ``semiinfinite planar dielectric'' will mean a finite
\notPR dielectric object whose 
\notPR surface curvature is much smaller than \( \kappa \), whose interior
\notPR contains no free charges, and whose volume is large enough that any
\notPR interior electric field would be prohibitively expensive in energy. As
\notPR two such objects approach, the gap \( \ell \) between them decreases
\notPR but the total volume occupied by solution doesn't change; this is why the
\notPR constant we subtracted from \eref{edfFfull} really is a constant.}
\yesPR{~\cite{fnc}}
bearing opposite
surface charge, thus changing the solution region from a half-space to
a planar slab of thickness \( \ell
\). Let
us suppose that a surface with $\sigma_+>0$ approaches another surface
with $\sigma_-<0$.

Parsegian and Gingell studied this situation in the
linearized approximation  \cite{pars72a}, arguing as in \sref{spp}
that the surfaces attract \via{} counterion release until all of one
species of counterions in the gap (the ``minority'' species) has been
exhausted. If \( \sigma_+\not=|\sigma_-| \), a residual cloud of the
other (``majority'') species  remains in the gap and  the
system equilibrates at a finite gap spacing \( \ell_* \). Nevertheless,
the final state has less free energy per unit area than it did
originally; the difference is
the {\sl adhesion strength} \( W \).

We could compute \( W \) by again solving a boundary-value problem as
in \sref{ssbf.lt}, but
there is a shortcut. Suppose that \(\sigma_+<|\sigma_-| \), so
that the $+$ counterions are the ``majority'' species.
In mechanical equilibrium the hydrostatic
pressure pushing the walls together vanishes. The planar Poisson-Boltzmann
equation is a second-order ordinary differential equation, and so its solutions
form a two-parameter family. One integration constant is fixed by the Gauss-law
boundary condition on the negative wall, while in equilibrium the
other is fixed by the condition of vanishing pressure. Hence the
solution \( \psb(z) \) is exactly the same for two walls as it
is for the isolated negative wall; the only difference is that in the
former case we truncate the solution at \( z=\ell_* \), while in the
latter case \( z \) extends to infinity. The equilibrium gap spacing
\( \ell_* \) is then just the value of \( z \) at which Gauss's law
for the positive wall is satisfied: \( -(-\pd\psi
z)={\sigma_+\over\epsilon} \). Then \eref{edfsolDH} gives the
equilibrium spacing $\ell_*$ by
\(\ex{\kappa\ell_*}=|\sigma_-/\sigma_+| \) in the linearized
approximation. Note that indeed the right side of this formula is
positive and greater than unity, as it must be since \( \ell_*\ge0 \).

We now recall that the linearized
approximation retains only the boundary term of \eref{edfFfull}, so
\begin{eqnarray}
f\gap(\sigma_+,\sigma_-)&=&
{\kbt\over2e}\left[\sigma_-\psb(0)+\sigma_+\psb(\ell_*)\right]\nonumber\\
&=&{1\over2\kappa\epsilon}\bigl( (\sigma_-)^2-(\sigma_+)^2\bigr)
\ .\label{edffgapLPB}\end{eqnarray}
Repeating these steps for the opposite case where
$\sigma_+>|\sigma_-|$, we find that in general (\tfig{ffthick}{\it a})
\beq f\gap(\sigma_+,\sigma_-)=|f\self(\sigma_+)-f\self(\sigma_-)|
\ .\label{edffgap}\eeq
Remarkably, the simple combination formula \eref{edffgap} will
continue to hold in the full nonlinear Poisson-Boltzmann treatment of
\sref{ssbf.npbt} below%
\notPR .\footnote{Behrens and Borkovec have
\notPR independently used this fact to simplify the study of nonlinear PB
\notPR solutions \cite{behr99b}.} 
\yesPR{~\cite{fnd}.}
Formula \eref{edffgap} is certainly
reasonable: when $\sigma_+=|\sigma_-|$ all counterions get released, the
two surfaces coincide, and this was our reference state of zero
energy. Also, when we reverse the signs of all the charges the free
energy should not change; \eref{edffgap} has this property.

Finally we find the adhesion energy $W$ as \cite{nard98a}
\beq W=f\self(\sigma_+)+f\self(\sigma_-) -f\gap
(\sigma_+,\sigma_-) =
\min\{(\sigma_+)^2,(\sigma_-)^2\}/\epsilon\kappa
.\ \mbox{(linearized approximation)}
\label{eWDH}\eeq
Note that $W$ is completely independent of the majority charge
density, a property noted by Nardi {\it et al.}
In light of the physical picture in \sref{spp}, we can readily
interpret that fact: The total
counterion release is limited by the {\it smaller} of the two
counterion populations.

Since $W$ is always positive we find, as expected, that {\it
oppositely-charged dielectrics always attract} via the
counterion-release mechanism \cite{pars72a}. Of course this is not the behavior
we
were seeking to explain (see \sref{sintro}). We must now proceed to
generalize the above
arguments, incorporating the relevant differences between the above
system and the one studied in the experiments of \cite{ramo99a}.

\subsection{Thick membrane}\label{sstm.lt}
We just found that two oppositely-charged dielectrics attract, as
expected. But the experiments we are studying involve dielectric (latex)
spheres interacting not with other dielectrics, but with a bilayer
{\it membrane}. In this subsection we begin to incorporate the new
physics associated with this situation.  We first  study the interaction
of a dielectric of fixed charge density \( \sigma_-<0 \) with a
positively-charged, very thick, membrane,
recapitulating some results of Nardi {\it et al.} \cite{nard98a}.

The new physical feature of this situation is that the bilayer
membranes in the experiments are {\it fluid mixtures} of
positively-charged and neutral surfactants. This means that the charge
density $\sigma_+$ on the membrane is not a fixed number, but may vary
subject to \( \sigma_+>0 \) and the overall constraint that the total
membrane charge \( \int\dd A\,\sigma_+ \) is fixed.
Let \( \spav \) denote the average charge density, so that the total
membrane charge is \( A\spav \). In addition we will suppose that
the charge density cannot exceed a maximum of \( \smax=2e/a_0 \)
determined by the area per headgroup $a_0$ of the charged surfactants
in each of the two monolayers constituting the membrane,
and that \( |\sigma_-|>\smax \).

Throughout this paper we will adopt a highly simplified,
generic picture of membrane compositional changes, retaining only the
entropy of mixing of the two surfactant types. Thus we neglect other
entropic or enthalpic packing effects in the assumed membrane free
energy \( \fm \). Moreover, at first we
will for simplicity neglect the bilayer structure of the membrane;
later on we will use formul\ae{} appropriate to a bilayer. With these
simplifications \( \fm \) takes the form
\beq \fm ={2\over a_0}\kbt\left[ {\sigma_+ \over\smax}\ln{\sigma_+ \over\smax}
+\Bigl(1-{\sigma_+ \over\smax}\Bigr)\ln\Bigl(1-{\sigma_+ \over\smax}\Bigr)
\right]
\ .\label{edffm}\eeq

As discussed in \sref{spp}, we wish to explore the possibility of a
spontaneous partition of the membrane into two uniform regions, which
we will call zones ``a'' and ``n''. (Ultimately we hope to find that
``a'' is {\it a}dhering while ``n'' is {\it n}onadhering, but for the
moment these are arbitrary names.) The areas of the two zones are not known in
advance, but they must add up to the total area $A$, so we take them
to be \( \gamma A \) and \( (1-\gamma)A \) respectively.

The two zones
exchange one conserved quantity, namely membrane charge%
\notPR .\footnote{A second
\notPR conserved quantity, the total charge of the counterions, is not
\notPR independent but instead fixed by charge neutrality.} 
\yesPR{~\cite{fne}.}
Thus the system can divide into zones of charge
density \( \sigma_+^{(a )}\) and \( \sigma_+^{(n)} \), subject to
\beq \gamma\sigma_+^{(a )} + (1-\gamma)\sigma_+^{(n)} =\spav
\ .\label{egammaconstraint}\eeq
This separation
will be energetically advantageous if the corresponding total free
energy \( A\bigl(\gamma f(\sigma_+^{(a )}) + (1-\gamma)f(\sigma_+^{(n)})\bigr)
\)
is less than \( Af(\spav) \). Here the free energy density  \(
f(\sigma_+) \) is computed for a {\it uniform }
zone with fixed membrane charge density \( \sigma_+ \), minimizing
over all other variables.

The instability
just described will not occur if
\( f(\sigma_+) \) is a convex function, i.e.\ \( \dd^2f/\dd{\sigma_+}^2>0
\). If \(
\dd^2f/\dd{\sigma_+}^2 \) is negative anywhere within the allowed
region \( 0<\sigma_+<\smax \), we apply the
Maxwell construction from thermodynamics to the graph of $f$. This
involves drawing a straight line tangent to the graph and spanning the
region of concavity. Let the two points of tangency be located at \(
\sigma_+^{(a )}\) and \(
\sigma_+^{(n)} \). If the average membrane composition \( \spav \)
lies between these two values, then the uniform system will
be unstable to partitioning into two zones with compositions
\( \sigma_+^{(a)} \) and \( \sigma_+^{(n)} \).

In the case of a thick membrane, we have \(
f=f\gap(\sigma_+)+\fm(\sigma_+) \). Consulting \eref{edffgap},
\eref{edffselfDH}, and \eref{edffm}, we find that the first
(electrostatic) term is
destabilizing, while the second (entropic) term is stabilizing. For
future use we introduce two convenient abbreviations, one parameterizing the
relative strengths of the two terms, the other a dimensionless measure
of charge density:
\beq \beta\equiv2\hat na_0/\kappa=\kappa a_0/4\pi \lB
\ ,\qquad \sbar\equiv\sigma/\smax
\  .\label{edfbetasigmabar}\eeq
With these abbreviations we obtain
\[f={{\smax}^2\over2\epsilon\kappa}\left[ |(\smb)^2
-(\spb)^2| + \beta\bigl(\spb\ln\spb +
(1-\spb)\ln(1-\spb)\bigr)\right]\  .\ {\rm (linearized\ approximation)}
\]
Nardi {\it et al.} pointed out that
this function has an inflection point, giving a region of instability
(\tfig{ffthick}{\it b}),
when \( \beta<1/2 \). According to \eref{edfbetasigmabar}, this means
that either the maximum charge density \( e/a_0 \) must be large, or
else the salt concentration \( \nhat \) very small.
Substituting some typical values for the charge per
headgroup \( a_0=0.5\,\)nm and salt concentration $\hat
n=1\,$mM$=0.0006\,$nm$^{-3}$ gives $\beta=0.006$, well into the regime of
instability.

\newcommand{\captwo}{{\it(a)}~Sketch of the electrostatic part of the
free energy ($f\gap$, formula \eref{edffgap}), for a thick, mixed,
membrane approaching a dielectric. We show the case where
$|\sigma_-|>\smax$. \hfil\break {\it(b)}~Sketch of the
total free
energy ($f\gap+\fm$, including formula \eref{edffm}), illustrating the
Maxwell construction. The features of the curve have been exaggerated
for clarity. The membrane will partition into a highly
attractive region with charge density  \( \sigma_+^{(a)} \) and a
somewhat-attractive region with  \(\sigma_+^{(n)} \).}

\notPR\ifigure{ffthick}{\captwo}{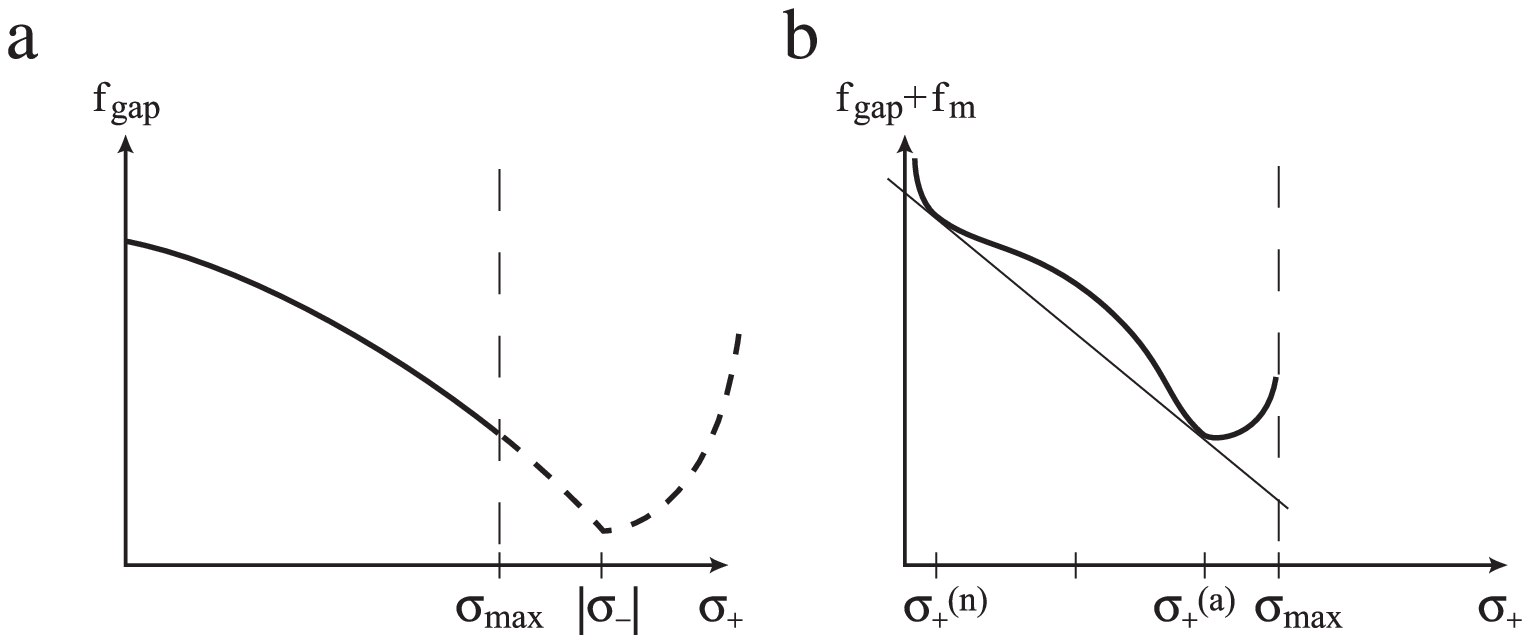}2

Though we have found an instability, two remarks limit its interest.
First, we have insisted that \( 0<\sigma_+<\smax \), so of course the
charge densities \( \sigma_+^{(a)} \) and \( \sigma_+^{(n)} \) on our
two zones are both positive: both zones are adhesive, unlike the
experimental phenomenon we are trying to explain.

Moreover, we found no instability at all unless the charge density
\( \smax \) is quite large (recall also that  \( |\sigma_-| \) is
assumed to be even greater than this). But at such large charges our
linearized approximation breaks down!  Our calculation becomes
inconsistent just as it gets interesting. Much of this paper is
dedicated to correcting this deficiency. We ask the reader to
suspend disbelief momentarily while we implement the physical picture
sketched in \sref{spp} in the linearized theory, where the formul\ae{}
are simple. Our claim is that the physical
picture is robust and holds beyond this inadequate mathematical
framework; we will support this claim by improving the
calculation in \sref{snpbt}--\ref{seic}.

\subsection{Thin, permeable membrane}\label{sstpm.lt}
The previous subsection found that a highly-charged, thick, membrane
can partition into a zone of strong adhesion and a second zone of
weaker adhesion. In this subsection we will introduce another element
of realism by accounting for the interior charges in the vesicle. To
highlight the key role of the membrane as a barrier to counterions, we
will first study the simpler case of a thin membrane {\it permeable}
to ions, finding uniform attraction. This sets the stage
for the more interesting case of an {\it im}permeable membrane in
\sref{sstim.lt} below.

Thus the new feature introduced in this subsection is that a membrane
separates the world into two compartments, with electrolyte solution
on each side (\tfig{fgapgeom}a).  We continue to neglect the internal
structure of the membrane, treating it as a single thin sheet of
charge; in \sref{ssftbm.npbt} below we will improve the analysis to
include the bilayer structure and finite internal capacitance of real
membranes.

\newcommand{\capthree}{{\it(a)}~Schematic for the approach of a
dielectric to a thin, permeable, membrane.
Negative values of $z$ in the text refer to the interior of the
vesicle.\hfil\break
{\it(b)}~Sketch of the free energy density $\fb$ as a function of \( \sigma\one
\),
holding \( \sigma_+ \) fixed.}

\notPR\ifigure{fgapgeom}{\capthree}{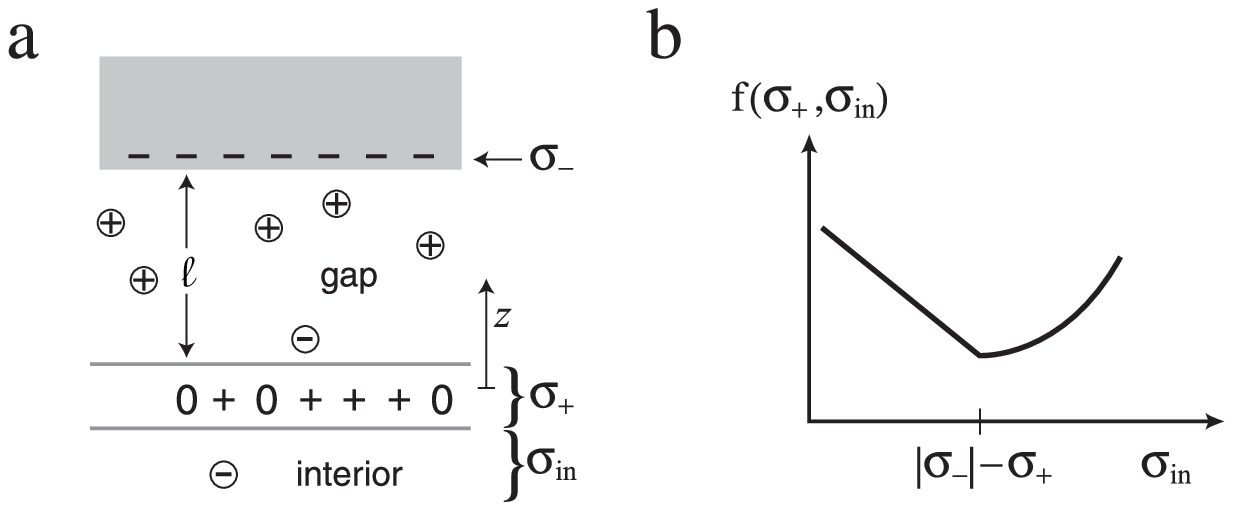}{2}

To organize the calculation we first note that once again there is
only one independent conserved quantity exchanged laterally between zones on
the
membrane, namely \( \sigma_+ \). Let
\beq \sigma\one\equiv\int_{-\infty}^0\dd z\,e(n_+(z)-n_-(z))
\  .\eeq
be the areal density of mobile interior counterions. Note that unlike
\( \sigma_+ \), which must be positive and less than \(\smax\), the
interior density \(\sigma\one\) can in principle have any sign and magnitude.
We will hold \(
\sigma\one \) fixed while optimizing over the gap spacing \( \ell \)
as in \sref{sstd.lt} above. Since
in this subsection we are assuming a permeable membrane, we then
minimize over \( \sigma\one \) as well to obtain \( f(\sigma_+) \).
We will suppress explicit mention of the dependence on the dielectric
charge $\sigma_-$, because  $\sigma_-$ is fixed.

The free energy density $f$ can be regarded as an interior term from
\eref{edffselfDH}, plus a gap term, \( f\gap \) from
\eref{edffgap}, plus the membrane free energy \( \fm \) from \eref{edffm}.
The interior term is \( f\self(-\sigma\one)=(\sigma\one)^2/2\epsilon\kappa \).
The gap sees opposing charge densities of \( \sigma_- \) and
\( (\sigma_++\sigma\one) \), so \eref{edffgap} gives
\( f\gap=|{\sigma_-}^2-(\sigma_++\sigma\one)^2|/2\epsilon\kappa\) and
the equilibrium spacing is
\( \ell_*(\sigma\one,\sigma_+)={1\over\kappa}\ln{|\sigma_-|\over
\sigma_++\sigma\one} \).

The
membrane free energy \( \fm \) is independent of \( \sigma\one \),
so minimizing over \( \sigma\one \)  gives (\tfig{fgapgeom}b) \( \sigma_{\rm
in,*}=|\sigma_-|-\sigma_+ \),
a positive value corresponding to
\( \ell_*=0 \): the membrane comes into tight contact with the
dielectric. Evaluating the free energy at this point gives
\beq
f(\sigma_+)={{\smax}^2\over2\epsilon\kappa}\left[
(|\smb|-\spb)^2 
+ \beta\left(\spb\ln\spb
+\bigl(1-\spb\bigr)\ln\bigl(1-\spb\bigr)\right)
\right]
\  .\label{eftpm}\eeq

Computing the second derivative we see that this time every term of
$f$ is separately convex: there is no
instability. Our result is physically reasonable. As the positive membrane
approaches the dielectric, the latter's negative (minority) counterions and
some of the
positive (majority) counterions get released to the exterior. Since we have
assumed the membrane is permeable, the remaining positive counterions
pass through it, where many more pair up with interior negative ions
from the membrane and get released to the interior. Since no
counterions need to remain in the gap, we get tight contact between
membrane and dielectric, which in effect become a single object of
reduced charge density \( \sigma_+-|\sigma_-| \). The electrostatic
self-energy of this composite object is a convex function, entropy never
favors phase separation,  and so there is no instability.

\subsection{Thin, impermeable membrane}\label{sstim.lt}
Previous subsections have shown that  charge mobility alone can lead
to an instability, but not to charge reversal (\sref{sstm.lt}), and that
introducing a coupled interior compartment alone does not even lead to
instability (\sref{sstpm.lt}).  Surprisingly, in this section and
\sref{sstim.npbt} below we will
find that combining these two unpromising ingredients with the
hypothesis of a membrane impermeable to ions {\it can} lead to a
charge-reversing instability%
\notPR .\footnote{We need not
\notPR assume the membrane to be impermeable to {\it water}; because the bulk salt
\notPR concentration is assumed the same on both sides, there will be
\notPR no net osmotic flow.}
\yesPR{~\cite{fnh}.}
The key observation is that an
impermeable membrane has {\it two}  conserved quantities independently
exchanged between  zones: the membrane charge \( \spb \)
and the interior counterion charge density \( \sob \)%
\notPR .\footnote{A third
\notPR exchanged quantity, the net counterion charge density {\it outside} the
\notPR membrane, is then fixed by charge neutrality:
\notPR $\sigma_0=-(\sigma_1+\sigma_+ +\sigma_-)$. Similarly the density of
\notPR neutral surfactants in the membrane is not independent, being
\notPR given by $(2/a_0)-(\sigma_+/e)$.  The numbers of individual
\notPR counterions of each species are {\it not}, however, conserved, since neutral
\notPR $\pm$ pairs can be exchanged with large reservoirs (the bulk solution
\notPR inside and outside the vesicle) without macroscopic charge separation.}
\yesPR{~\cite{fnf}.}
In the previous subsection \( \sob \) could relax within a zone by passing
through the membrane, and so we simply optimized $f$ over it before
applying the Maxwell construction. For an impermeable membrane we must
instead apply the Maxwell construction to both \( \spb \) and \(
\sob \) jointly.

The geometry is the same as \tfig{fgapgeom}a. It will shorten some
formul\ae{} to define the total charge density
\beq\stb\equiv\spb+\sob\label{edfsbt}\ .\eeq
The free energy density
is the same as in \sref{sstpm.lt}, but this time we need a more
general formulation than \eref{eftpm}, since we are not simply
evaluating at the optimal value of \(\sob\). In fact, there are three
physically separate cases we must distinguish:\hfil\break
{\it i)} The membrane plus its trapped interior
counterions may have greater charge density than the dielectric:
\(\stb>|\smb|\).\hfil\break
{\it ii)} The membrane  plus its trapped
counterions may have lower charge density than the dielectric, but
still be positive: \(0<\stb<|\smb|\).\hfil\break
{\it iii)} The trapped counterions may overwhelm and
effectively reverse the charge of the membrane: \(\stb<0\). This is
the charge-reversal we seek. In this case the equilibrium distance
between the membrane and a negative dielectric is infinity; the
membrane actually repels incoming negative objects.

The total free energy density in each of these cases (and still in the
linearized approximation) now reads
\begin{eqnarray}
f(\spb,\stb)&&={{\smax}^2\over 2\kappa\epsilon}\Bigl[
(\stb-\spb)^2
+\left\{
\begin{array}{l@{\quad{\rm if}\ }l}
|(\stb)^2-(\smb)^2|&\stb>0\\
(\stb)^2+(\smb)^2&\stb<0\end{array}\right\} \nonumber\\
+&&\beta\bigl(\spb\ln\spb+(1-\spb)\ln(1-\spb)\bigr)\Bigr]\ .\quad {\rm
(linearized\ approximation)}
\label{efthinDH}\end{eqnarray}

The function $f$ defines a surface over the \((\spb,\stb)\)-plane. If
this surface is everywhere convex-down then there is no
instability. If not, then it may be possible to bring a
straight line up to the surface from below, touching it at two points
of tangency but lower than the surface at some point
\((\spavb,\stavb)\) lying between those tangency points. In this
case a homogeneous system with average composition
\((\spavb,\stavb)\) will be able to reduce its free energy by
partitioning into zones whose compositions are given by the two points
of tangency.

We will assume that initially, when the membrane vesicle was formed and no
dielectric spheres were present, the ions were in equilibrium across
the membrane, so that half of them got trapped inside:
\(\stavb=\half\spavb\). We will also for illustration generally take the
membrane
composition to be half charged and half neutral surfactants, so that
the mole fraction
\(\spavb=\half\). Finally we assume the approaching dielectric to have
greater charge density than the maximum possible value for the
membrane. For illustration we take \(\smb=-3/2\). Summarizing, we will
consider the illustrative case
\[\spavb=\half\ ,\qquad\stavb=\half\spavb\ ,\qquad\smb=-3/2\ .\qquad{\rm
(illustrative\ case)}\]
In general there may be many lines in the \((\spb,\stb)\)-plane, all
passing through the  point \((\spavb,\half\spavb)\) and all
exhibiting the instability. In this case we must examine all the lines
and choose the one which gives the absolute minimum in free energy. To
do this systematically, we label the lines by the real number $p$ and
write each parametrically as
\beq \bigl(\spb,\stb\bigr)=
\bigl(s,\lfr{1-p}2\spavb+\lfr p2 s\bigr)\ ,\qquad 0<s<1
\ .\label{edfpconstruction}\eeq

In principle one could now plot $f$ from \eref{efthinDH} along the
family of lines defined by \eref{edfpconstruction}, find the points of
tangency, optimize over $p$, and finally obtain the sought instability
and the charge densities \( \stb^{(a)} \) and \( \stb^{(n)} \)
in the two zones as points of tangency, as described in \sref{sstm.lt} above.
If one of
these (conventionally \( \stb^{(n)} \)) proves to be negative, then we
conclude that the membrane exhibits a charge-reversal instability, as
was to be shown. In fact these steps are now rather easy to
complete. But we have already remarked that the linearized theory is
not accurate in the regime of high charge densities of interest to us.
Accordingly we will now  improve our theory by solving the full nonlinear
Poisson-Boltzmann equation, then carry out the steps just described.

\section{Nonlinear Poisson--Boltzmann Theory}\label{snpbt}
\subsection{Basic formul\ae}\label{ssbf.npbt}
We introduce the useful new variable
\beq \zeta\equiv\ex{\kappa z}\ .\eeq
It will also
be convenient to define another nondimensional form of the charge
density by
\beq \sbb\equiv\sigma\kappa/2\hat ne=2\sbar/\beta\label{edfsbb}\ .\eeq
and a nondimensional form of the free energy density by
\beq \fb={\kappa\over \nhat\kbt}f\label{edffbar}\ .\eeq

The
general solution to the Poisson-Boltzmann equation can be written in
terms of elliptic functions (see for example
\cite{behr99b}). Fortunately, however, we need only the zero-pressure
solutions, corresponding to two walls which are free to adopt their
equilibrium spacing $\ell_*$, and these solutions consist of
elementary functions \cite{verveybook,russelbook}:
\begin{equation} \psb_\pm=\pm2\ln{\zeta+1\over
\zeta-1}\label{ePBsols}
\ .\label{ePBsol}\end{equation}
At $\zeta\to\infty$ we then have $\psb_\pm\to\pm4/\zeta=\pm4\ex{-\kappa
z}$, precisely the weak-field solution we found in \sref{ssbf.lt}.

The Poisson-Boltzmann equation is second-order, and so its general
solution has two integration constants. We have fixed one of these by
restricting to the zero-pressure case. The other one enters
\eref{ePBsols} rather trivially, due to the translation invariance of
the PB equation: in \eref{ePBsols} we are free to shift $z$, or equivalently
multiply $\zeta$ by an arbitrary constant, thus obtaining a
one-parameter family of zero-pressure solutions. In practice we will
use \eref{ePBsols} in the unshifted forms given above, but select the
region \( \zeta_1<\zeta<\zeta_2 \) to enforce Gauss's law at each of
the two charged surfaces.

For example, for an isolated surface of charge density $\sigma_+>0$ we
choose the solution $\psb_+$ with one limit at infinity and the other
at $\zeta_+$, which we choose by requiring
\[ {\sigma_+\over\epsilon}=-{\kbt\over e}{\dd\over\dd
z}\psb_+\Bigl|_{z_+}\ . \]
or
\beq \zeta_+={2\over\spbb}\left(1+\sqrt{1+\spbb^2/4}\right)
\ .\label{edfzetaplus}\eeq
The free
energy formula analogous to \eref{edffselfDH} is then obtained by
substituting \eref{ePBsols} into \eref{edfFfull}, to get
\beq
\fb\self=
{{-8\,\left( 2 - \zeta_+ \,\ln ({{1 + \zeta_+ }\over {-1 + \zeta_+ }})
\right)}
    \over {-1 + {{\zeta_+ }^2}}}
+2\spbb\ln{\zeta_++1\over\zeta_+-1}\qquad{\rm (nonlinear\ theory)}
\ .\label{edffselfPB}\eeq
Two surfaces of the same sign charge will repel to infinite separation, so
we use this formula for each one separately. For a negatively-charged
surface we simply replace $\spbb$ by $|\sbb_-|$ in
\eref{edfzetaplus}.

\newcommand{\capfour}{The self-energy density of a charged surface $\bar
f\self=\kappa
f\self/\kbt\nhat$ as a function of the dimensionless charge density
$\sbb$. Top curve:
Linearized approximation, eqn.~\eref{edffselfDH}. Bottom curve:
Poisson-Boltzmann solution, eqn.~\eref{edffselfPB}. $f\self$ is
positive in our conventions; see \sref{ssbf.lt}.}

\notPR\ifigure{ffselfPBDH}{\capfour}{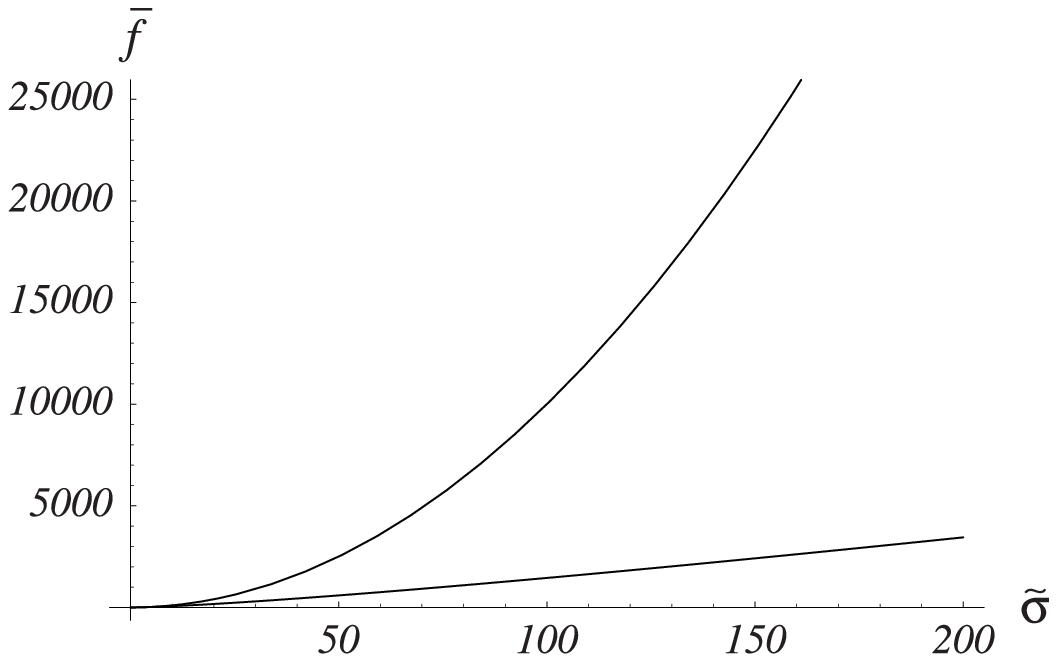}{2}

It is instructive to compare \eref{edffselfPB} to the corresponding
formula in the linearized approximation, formula \eref{edffselfDH}
(\tfig{ffselfPBDH}). While the two formul\ae{} agree at low charge
density, the linearized formula overestimates the free energy by
almost an order of magnitude at the high charge densities of interest
to us. The
nonlinear PB equation also predicts a narrower cloud of counterions than
the linearized approximation at any given charge density.

For two oppositely charged surfaces at their equilibrium spacing we
can generalize the argument given in \sref{sstd.lt} above, again
obtaining \eref{edffgap}. Again suppose first that
$\sigma_+<|\sigma_-|$, and so $\zeta_+>\zeta_-$. By the same logic as
in \sref{sstd.lt}, the potential $\psb$ in the gap is just the same as
that of an isolated surface of charge density $\sigma_-$, but
truncated at some finite $\zeta_+$.

Let us abbreviate the local integrand in \eref{edfFfull} by
$\Phi=\psb\sinh\psb-2\cosh\psb+2$. This is the same for either of the
two solutions $\psb_\pm$. We thus get
\[\fb\self(\sigma_\pm)=\pm{\kappa\sigma_\pm\over2e\nhat}\psb_+(\zeta_\pm)
+\int_{\zeta_\pm}^\infty\dd\zeta\,\Phi\]
and
\[\fb\gap(\sigma_+,\sigma_-)={\kappa\over2e\nhat}\bigl(
\sigma_-\psb(\zeta_-) +\sigma_+\psb(\zeta_+)
\bigr)+\int_{\zeta_-}^{\zeta_+}\dd\zeta \ \Phi \ .\]
But the last expression just equals
$\fb\self(\sigma_-)-\fb\self(\sigma_+)$. Repeating for the opposite
case $\sigma_+>|\sigma_-|$, we get the desired combination formula
\eref{edffgap}.

\subsection{Thin, impermeable membrane}\label{sstim.npbt}
Proceeding now as in \sref{sstim.lt}, we combine \eref{edfsbb},
\eref{edffbar}, \eref{edfzetaplus}, \eref{edffselfPB},
\eref{edffgap}, \eref{edffm}, and \eref{edfbetasigmabar} to obtain the
analog of the linearized formula \eref{efthinDH}: the nondimensional free
energy density
of the membrane+dielectric system at its equilibrium spacing $\ell_*$,
as a function of the local membrane charge density $\sigma_+$ and the
net charge density $\sigma_{\rm t}$ of counterions trapped inside the
membrane vesicle, is now
\begin{eqnarray} \fb(\spb,\stb)&=&\fb\self(\stb-\spb)+\left\{
\begin{array}{l@{\quad{\rm if}\ }l}
|\fb\self(\stb)-\fb\self(\smb)| & \stb>0\\
\fb\self(\stb)+\fb\self(\smb) &\stb<0\end{array}\right\} \nonumber\\
&+&{4\over\beta}\bigl(\spb\ln\spb+(1-\spb)\ln(1-\spb)\bigr)
\ .\label{efbartimPB}\end{eqnarray}
Here $\stb=\spb+\sob$ as before and $\fb\self(\sbar)$ is the
function defined by \eref{edfzetaplus}, \eref{edffselfPB} evaluated at
$\sbb=2\sbar/\beta$. Thus at low charge densities the first two terms
of \eref{efbartimPB} contain factors of $1/\beta^2$, and so dominate
the last (mixing-entropy) term, just as in \eref{efthinDH}.

\newcommand{\capfive}{Free energy density $\fb(\spb,\stb)$  for a
thin, impermeable, membrane. For
easier visualization we have inverted the figure, rescaled, and added
a linear function, plotting \offsetthreed{} instead of $f$.
The solid curve
is the locus of points where $\stb=0$; points to the left of this curve
represent charge-reversed states. The heavy dot is the point
$(\spavb,\stavb)=(1/2,1/4)$ representing the average membrane
composition chosen for our illustrative calculation.
The two hills in the graph imply, {\it via} the Maxwell construction, that
the system's ground state consists of two coexisting zones. Since
furthermore the hills straddle the solid curve, one of the zones is
charge-reversed. [Adapted by  permission from \cite{aran99a}. \copyright{} 1999
American Association for the
Advancement of Science.]}

\notPR\ifigure{ffbartim3D}{\protect\capfive}{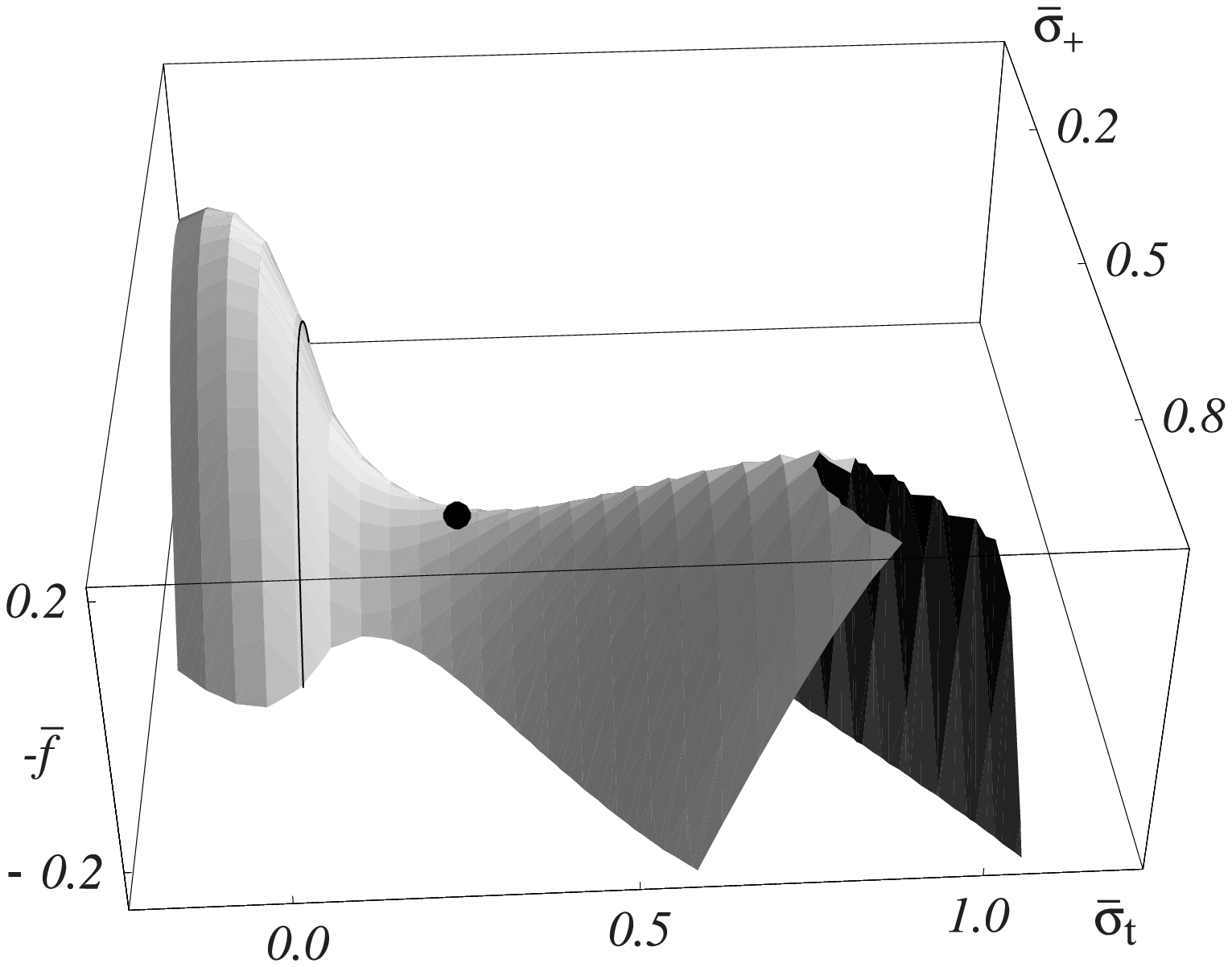}{3}

We can now carry out the program outlined in \sref{sstim.lt} for the
illustrative parameter values $\nhat=1\,$mM, $a_0=0.5\,$nm$^2$,
$\beta=0.006$, $\smb=-3/2$, and $\spavb=1/2$ discussed
earlier. \tfig{ffbartim3D} shows the surface defined by
\eref{efbartimPB}. For clarity we have shown $-\fb$ instead of $\fb$, so
that thermodynamic stability would correspond to an inverted bowl
shape. We have also tilted the graph by adding a convenient linear
function to $-\fb$, to highlight the saddle shape. The linear
function was selected by trial and error. Adding it does not change
the points of tangency between the surface and a straight line.

The graph clearly displays the instability we were seeking. Moreover,
one of the two hills on the surface clearly lies to the left of the
line of charge reversal, $\{\stb=0$\}. To make this qualitative
observation precise, we must now evaluate \eref{efbartimPB} along the
family of lines specified by \eref{edfpconstruction}, perform the
Maxwell construction on each line, and choose the value of $p$ whose
tangent line has the lowest value of $\fb$ at the point
$\spavb$. \tfig{fthreecurves} shows the result of this
analysis for the illustrative values  $p=2.4$ and 3.8, and the optimal
value $p=2.9$.

\newcommand{\capsix}{Three slices through \tfig{ffbartim3D} along lines
passing through the average-composition point $(\spb,\stb)=(1/2,1/4)$. The
total
charge density seen from outside the membrane is taken to be
$\stb=((1-p)\spavb+p\spb)/2$ for various values of $p$. Unlike
\tfig{ffbartim3D} we have not inverted this graph. Two illustrative
slices (gray curves) show $p=2.4$ and 3.8. The black curve, with $p=2.9$,
gives the most advantageous mixed
state, since its tangent line intersects $\spb=\spavb=1/2$ at the
lowest value of $\fb$; hence $p_*\approx2.9$. The dashed line shows coexistence
between
an adhesion zone with $\spb^{(a)}=0.95$, covering a fraction $\gamma_*=0.36$
of the vesicle, and a nonadhesive zone with $\spb^{(n)}=0.25$. The heavy dot
shows the point of charge reversal, where $\stb=0$. Since the points of
tangency lie on opposite sides of this dot, the nonadhesive zone
indeed presents net negative charge to the outside world. Again the curves have
been tilted for
viewing by plotting \offsetcurves.}

\notPR\ifigure{fthreecurves}{\capsix}{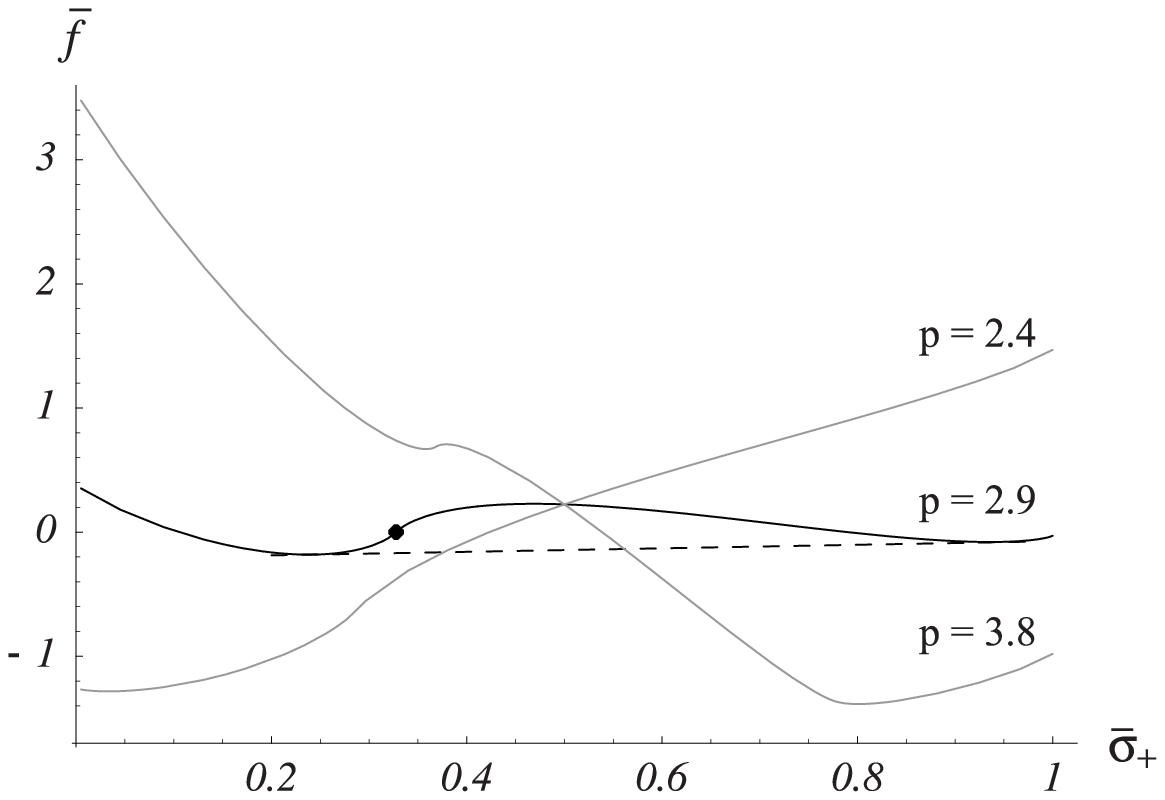}{2.7}
\notPR

The figure shows coexistence between
a  zone with $\spb^{(a)}=0.95$, and another zone with
$\spb^{(n)}=0.25$. The latter
zone thus presents total charge density $\stb=-0.11$ to the outside of
the vesicle. Since this is negative, this zone is charge-reversed and
deserves its name as a ``nonadhesive'' zone. Indeed the effect is
large: $\stb$ is $-45$\% as great as the charge $\spavb/2=1/4$ presented to
the outside world when there are no adhering dielectrics spheres.
Recalling that $\gamma\spb^{(a)}+
(1-\gamma)\spb^{(n)}=\spavb$, we find that the adhesion zone covers 36\%
of the vesicle. These results were announced in \cite{aran99a}.

\subsection{Finite thickness, bilayer membrane}\label{ssftbm.npbt}
While the above results are encouraging, and show the mathematical
possibility of a charge-reversal instability, our model needs
considerable refinement before we can take its results seriously. In
this subsection we begin this task by acknowledging the bilayer
character of the membrane and its finite capacitance, both neglected
up to this point. The results in this section were also announced in
\cite{aran99a}.

Instead of idealizing the membrane as a thin sheet of charge density
$\sigma_+$, we now regard it as two sheets of charge density
$u\sigma_+$ and $(1-u)\sigma_+$ representing the charged headgroups of
the inner and outer surfactant layers respectively (see
\tfig{fsetup}d). These two layers
of charge are separated by a dielectric layer of thickness $t$ and
dielectric constant $\epsilon_{\rm m}$, creating a capacitor of
capacitance $c=\epsilon_{\rm m}/t$ per area. We will estimate $c$
using the value 0.01~pF/$\mu\rm m^2$ typical for artificial bilayer
membranes \cite{genn89a} and make the useful abbreviation
\beq\tau\equiv t\kappa\epsilon/\epsilon_{\rm m}=
\kappa\epsilon/c
\ .\label{edftau}\eeq
Then $\tau\approx7$ at salt concentration $\nhat=1\,$mM, or more
generally $\tau\approx7\sqrt{\nhat/{\rm mM}}$.

The free energy formula \eref{efbartimPB} used in \sref{sstim.npbt}
needs only two simple modifications:\hfil\break
1)~Since the membrane still presents charge density
$\sigma_+-\st$ to the interior solution and $\st$ to the exterior
(\tfig{fsetup}), the terms involving $\fb\self$ are unchanged. Now,
however, when $\sigma\one+u\sigma_+$ is nonzero (or equivalently
$\st+(u-1)\sigma_+\not=0$), there will be a nonzero electric field in
the membrane's interior, with a capacitive energy cost per area of
$\bigl( \st+(u-1)\sigma_+\bigr)^2/2c$.\hfil\break
2)~The membrane now consists of {\it two} fluid monolayers of mixed charged
and neutral surfactants. Each monolayer has a maximal density
$\half\sigma\max =e/a_0$, attained when the density of neutrals is
zero. Accordingly we replace the mixing entropy $\fm$ from
\eref{edffm} by
\[ {\kbt\over a_0}\left[
{u\sigma_+\over \sigma\max/2}\ln{u\sigma_+\over
\sigma\max/2}+\cdots\right] \ .\]

Casting everything into the nondimensional forms defined above, we see
that we must add to the formula \eref{efbartimPB} for $\fb$ the
expression
\begin{eqnarray}
{4\tau\over\beta^2}\bigl(\stb+(u-1)\spb\bigr)^2&&+
{2\over\beta}\Bigl[2u\spb\ln(2u\spb)+(1-2u\spb)\ln(1-2u\spb)\nonumber\\
+2(1-u)&&\spb\ln 2(1-u)\spb+(1-2(1-u)\spb)\ln(1-2(1-u)\spb)\Bigr]
.\label{eaddbilayer}\end{eqnarray}
To use our formul\ae{} we add \eref{eaddbilayer} to \eref{efbartimPB}
and again hold fixed the two conserved
quantities $\spb$ and $\stb$. As
before we must optimize over all other variables, in this case just
$u$, before performing the Maxwell construction as in
\sref{sstim.npbt}. We can simply optimize \eref{eaddbilayer} over $u$,
since $u$ does not enter \eref{efbartimPB}. However, this optimization
is subject to the four inequalities which $u$ must obey:
\[0<u\spb<1/2\ ,\quad 0<(1-u)\spb<1/2\ .\]

A graphical analysis similar to the one shown in \tfig{fthreecurves}
now gives (\tfig{fthickmemb}) that the best line through
$(\spb,\stb)=(1/2,1/4)$ has $p\approx2$
(see \eref{edfpconstruction}), with points of tangency at
$\spb^{(a)}=0.247$, and $\spb^{(n)}=0.65$. Proceeding as in
\sref{sstim.npbt} gives coverage $\gamma_*=63$\% at equilibrium and
$\stb=-0.003$, or about $-1.2$\% of the charge density presented to the
outside world when no dielectric spheres are present.

\newcommand{\capseven}{Free energy density along the optimal line
$p\approx2$ for a finite-thickness, bilayer membrane. The long tick on
the abscissa is at $\spavb$, the illustrative value for average bilayer
concentration studied in the text. The point of
charge reversal is $\spb=0.25$, which is the second point from the left where
the graph intercepts the abscissa. Since the tangent point is slightly
to the left of this, at $\spb^{(n)}=0.247$, we again find a slight
charge reversal. Again the graph has been scaled and shifted to bring
out its structure:  we have plotted the quantity \offsetthick.}

\notPR\ifigure{fthickmemb}{\capseven}{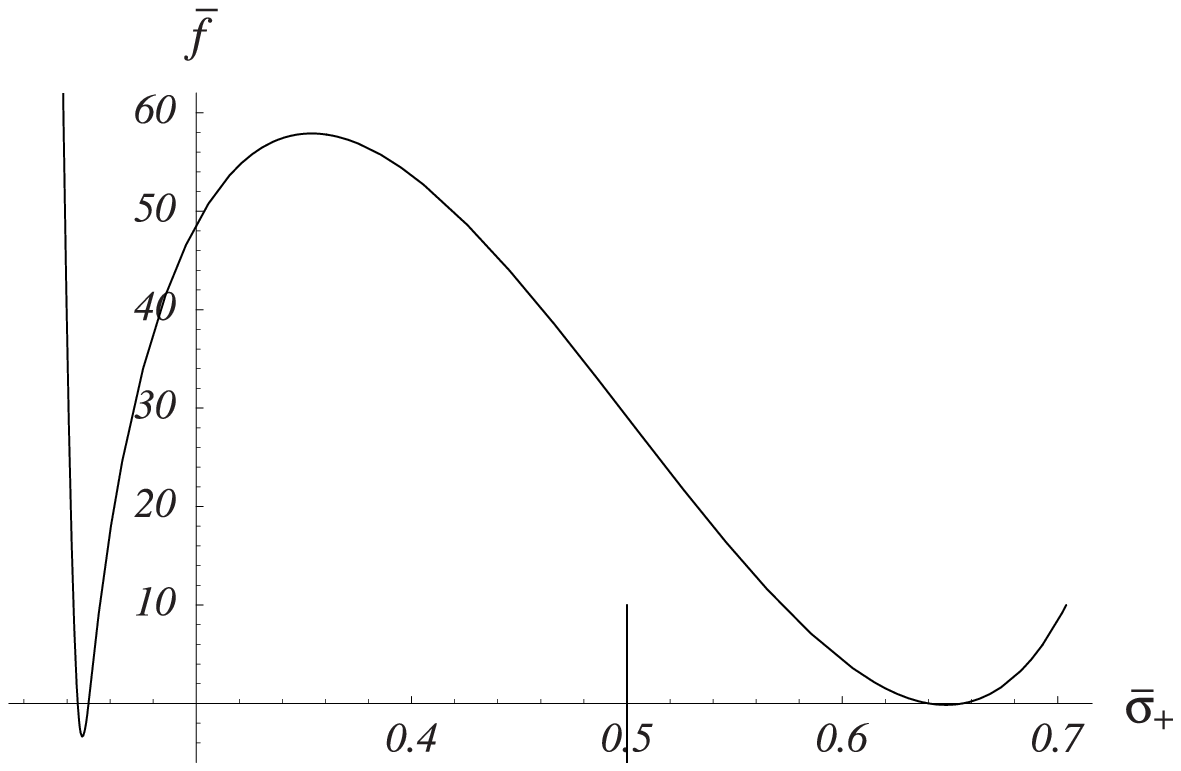}{2.5}

We can readily understand the qualitative features of these
results. Since $\tau$ is large, electric fields inside the membrane
are energetically costly and the two sides of the membrane are nearly
independent. Thus the interior ion charge density $\sigma\one$ remains
nearly uniform, and hence nearly equal to $-\spav/2$, and similarly the inner
monolayer charge density $u\sigma_+\approx\spav/2$. Then the total
charge density $\stb\approx(1-u)\spb\approx{1-u\over2u}\spavb$ (see
\tfig{fsetup}{\it d}), and \eref{edfpconstruction} requires that
either $u=1/2$ or $p\approx2$. The solution $u=1/2$ is unphysical; the
solution $p\approx2$ is just what we found numerically. To reverse its
charge, the membrane must allow electric 
fields in its interior (see
\tfig{fsetup}{\it d}); the high cost of doing this accounts for the
very sharp left-hand dip in \tfig{fthickmemb} compared to
\tfig{fthreecurves}, and for the greatly reduced degree of charge
reversal in the finite-thickness case.
The charge-reversal effect diminishes for larger
values of $\tau$, as we predicted in \sref{spp}. According to
\eref{edftau} this means the effect will disappear either for thick
membranes or at large enough ion strength $\nhat$. Numerically we find
the critical value to be about
$20\,$mM, roughly as seen in the experiments of \cite{aran99a,ramo99a}.

Though the charge-reversal effect seems small, it is enough to cause
the rejection of additional negative dielectric spheres. To estimate
the magnitude of this effect,
consider what is needed to increase $\gamma$ from its equilibrium
value $\gamma_*$ to $\gamma_*+\delta$. To do this we must choose new
values of $\spb^{(a)}+\epsilon^{(a)}$ and $\spb^{(n)}+\epsilon^{(n)}$,
subject to the condition \eref{egammaconstraint}, which now reads
\[(\gamma_*+\delta)(\spb^{(a)}+\epsilon^{(a)})
+(1-\gamma_*-\delta)(\spb^{(n)}+\epsilon^{(n)})=\spavb\ .\]
We then minimize the total free energy $F$ over $\epsilon^{(a)}$ and
$\epsilon^{(n)}$ subject to this constraint, finding that the increase
in $F$ when we force a nonequilibrium value of $\gamma$ is
\[ \Delta
F={\kbt\nhat\over\kappa}{1\over2}(\spb^{(a)}-\spb^{(n)})^2\left[
{1-\gamma_*\over \fb^{\prime\prime}_{(n)}}+{\gamma_*\over\fb^{\prime\prime}_{(a)}}
\right]\inv\,A\delta^2 \ .\]
In this expression $\fb^{\prime\prime}_{(a)}$ denotes
${\dd^2\fb\over\dd{\spb}^2}\Bigl|_{\spb^{(a)}}$, \etc, and $A$ is the
total membrane area.  Bringing an additional 1$\,\mu$m$^2$ of negative
dielectric into contact with a vesicle of area $4\pi(10\,\mu{\rm
m})^2$ gives $\delta=0.00083$. Evaluating numerically then gives
$\Delta F\approx 3000\kbt$, a huge barrier to adhesion, and similarly
if we pull away 1$\,\mu$m$^2$ of adhering dielectric.

\section{Effects of Ion Correlations}\label{seic}
The charge density near a highly-charged surface can become so great
as to invalidate the mean-field theory we have used so far. The
resulting changes in the force between two plates have been the object
of intense study since the discovery that the total force can become
attractive for two {\it like}-charged plates, in the presence of
multivalent counterions \cite{guld84a,kjel84a}. The situation we will
need to study will be much simpler than that one, for two
reasons. First, we are only interested in the free energy density at
{\it zero} force (\ie, equilibrium). Secondly, our effect has arisen
already at the level of mean-field theory. Since
we consider the case of {\it mono}valent
counterions only, \ie{} in the regimes of small to
moderate ion interactions, correlation effects will turn out to be a
modest correction to the main, mean-field, contribution. Thus we are
in the regime opposite to that recently studied in
refs.~\cite{rouz96a,pere99a}.

The criterion for mean-field theory to be valid is roughly that the
electrostatic potential energy of two ions at the mean ionic
separation be smaller than the thermal energy: $e^2n^{1/3}_\pm/4\pi
\epsilon<\kbt$, or equivalently that $n_\pm\lB^3<1$. Any isolated surface,
no matter how highly charged, will have a distance beyond which this
criterion is satisfied, and so our universal Poisson-Boltzmann solution
\eref{ePBsol} will be valid there. We will call the boundary of this
region $z=z_s$. We find $z_s$ using \eref{ePBsol} and \eref{edensity},
obtaining $z_s=0.28\,$nm for our illustrative case of ambient salt
concentration $\nhat=1\,$mM.

Thus any isolated planar surface has exactly the same
potential (equation \eref{ePBsol}) as any other, for $z>z_s$. The
charge density $\sigma_+$ enters only \via{} the location $z_+$ of the surface
in the coordinate $z$. Given a surface of
charge density $\sigma_+$, we compute $z_+=\kappa\inv\ln\zeta_+$,
where $\zeta_+$ is given by \eref{edfzetaplus}. If $z_+>z_s=0.28\,$nm, then
mean-field theory is everywhere accurate and there is no
correlated-ion cloud near the surface. In the opposite case, that
part of the ion cloud lying within the layer $z_+<z<z_s$ will have
nonnegligible correlations. Since $z_+$ is always positive, this layer
is never any thicker than a typical ion radius, and so may be treated
as a two-dimensional classical charged gas \cite{tots78a}. This
approach may be regarded as an approximation to other, more refined,
calculations (\eg{} refs.~\cite{kjel96a,atta88a}).

The effect of correlations will be to reduce the free energy density, as ions
can avoid each other, reducing their electrostatic self-energy. To
apply the results of Totsuji, originally derived for use in the study
of electrons adsorbed onto liquid helium \cite{tots78a}, we need to
know the two-dimensional density $m$ of counterions in the correlated
layer. $m$ simply equals
$\sigma_+/e$ minus the total density in the
uncorrelated region $z>z_s$. Again using \eref{ePBsol} and
\eref{edensity} in the latter region, we find
$m=(\sigma_+/e)-0.81\,{\rm nm}^{-2}$. If this quantity is negative
then there simply is no correlated layer and $m=0$. Defining the
plasma parameter as $\Gamma=\lB\sqrt{\pi m}$,
the correlation
energy density can then be represented by the interpolation formula
$E_c=m\kbt\Gamma(-1.07+{1\over2.2\Gamma+1.3} )$, which is approximately
valid over the range $0<\Gamma<5000$ \cite{tots78a}. \tfig{ftotsuji}
shows the resulting change in the free energy density, obtained from
the thermodynamic formula $f_c(m)=m\kbt \int_0^\Gamma
{\dd\Gamma\over\Gamma}{E_c\over m\kbt}$ for the correlation
contribution $f_c$ to the free energy density.

\newcommand{\capeight}{Correction to the self-energy density
$\fb\self$ from counterion correlations. The upper curve is the
Poisson--Boltzmann result (lower curve of \tfig{ffselfPBDH}); the lower curve
includes the correction $f_c$ discussed in the text.}

\notPR\ifigure{ftotsuji}{\capeight}{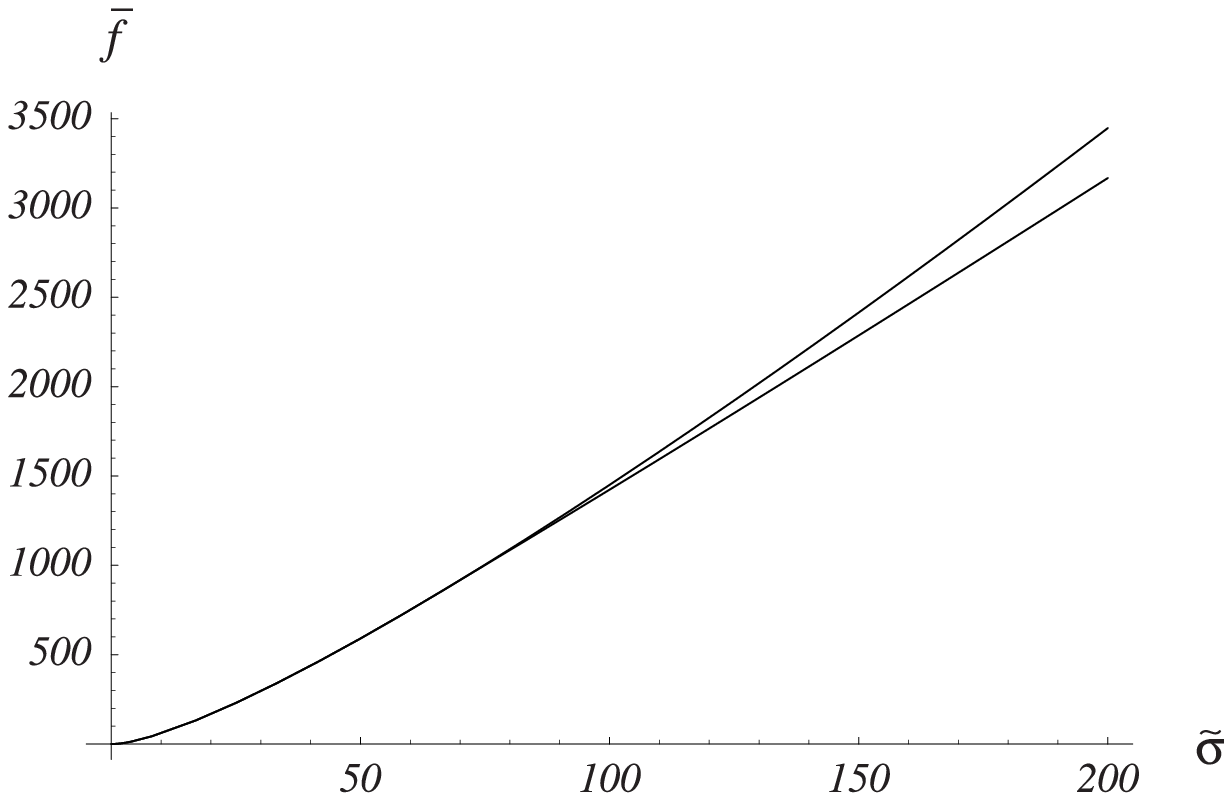}{2}

When two oppositely-charged surfaces face each other, we have seen how
the minority surface will be stripped of its counterions; the majority
surface may, however, have a correlated ion cloud, so we add $f_c(m)$
to its free energy density. Since in the situations of interest to us
the majority surface is always the dielectric sphere, and since $m$
depends only on the surface's charge (not on the presence of the other
surface), we find that {\sl the correlation correction to the free energy of
the negatively-charged dielectric surface is a constant}, and may
simply be dropped. We do need to include the correlation free energy
of the membrane's interior surface, but for a finite-thickness
membrane this too is nearly a constant, since as we have seen the
inner monolayer's charge density $u\sigma_+$ deviates only slightly
from $\spav/2$. Finally, in conditions of charge reversal the {\it
outer} monolayer becomes isolated and can have an ion cloud of its
own. Since as we have seen the degree of charge reversal is very
small, the density $m$ of this last ion cloud is very small and the
correlation contribution is negligible.

We have just outlined qualitatively why counterion correlations may be
expected to have little effect on the results given in
\sref{ssftbm.npbt}. Indeed, the numerically calculated graph analogous to
\tfig{fthickmemb}
is not appreciably different from that graph, and we do not display it
here.

\section{Discussion}\label{sd}
We have proposed a theoretical explanation for the phenomenon of
electrostatic adhesion saturation observed experimentally in
\cite{aran99a,ramo99a}. While the experimental system has not been
systematically explored yet, our model reproduces qualitatively the
surprising the phenomenon of charge reversal and
several salient experimental facts
\cite{aran99a,ramo99a}:\hfil\break
1) Adhesion saturation occurs only with {\it mixed} bilayer vesicles,
that is, at mole ratios $\spavb$ not too close to zero or
unity.\hfil\break
2) It occurs only under conditions of sufficiently low
salt.\hfil\break
3) The saturated state has a very definite number of adhering objects
($\gamma_*$ is fixed for each vesicle).

Our analysis has omitted many familiar colloidal-force effects. Many
of these are short-ranged (\eg{} solvation forces), weak compared to
electrostatic forces (\eg{} undulation repulsion), or rapidly
decreasing with distance (\eg{} van der Waals forces). In addition we
have neglected all finite ion-size effects. We believe that
our conclusions will be robust when such effects are introduced, in
part because the crucial physics of charge reversal involves the
immediate neighborhood of the left-hand dip in \tfig{fthickmemb},
namely, the separation between the charge-reversal point and the
tangency point. But the distance $\ell_*$ between the membrane and
dielectric diverges as we approach the charge-reversal point from the
right, so this physics is controlled by the long-distance behavior of
the forces. Certainly the exact location of the tangent point depends
on the right-hand part of \tfig{fthickmemb} as well, where our theory
is not reliable. But this dependence is small due to the
sharpness of the left-hand dip in the free energy density. Even if
the right-hand side of the graph differs from what we computed,
there should be a range of  membrane compositions $\spavb$ greater
than $\spb^{(n)}$ but  low enough to be in
the left part of the graph, and hence yielding the sort of zone separation
we have studied.

We have examined only {\it equilibrium} states. It is quite possible
that the experimental system of \cite{aran99a,ramo99a} is not in
equilibrium, \ie\ that the observed coverage $\gamma$ is less than
the equilibrium value $\gamma_*$ because the last one or two balls is
initially repelled by a finite free-energy barrier. But our goal was
to understand the surprising existence of {\it any} barrier,
not to predict a specific value for $\gamma_*$, which in any case
depends on the membrane composition%
\notPR .\footnote{Moreover, the observed ball
\notPR coverage does not directly give $\gamma$, since the degree of
\notPR each 
\notPR ball's coverage is not optically observable; see
\notPR \cite{ramo99a}.} 
\yesPR{~\cite{fng}.}

The analysis suggests a number of experimental tests of our mechanism.
A mixed vesicle adhering to a charged dielectric surface
\cite{nard98a} may provide a more controlled geometry than that of
\cite{ramo99a}; in this case adhesion saturation suggests the
possibility of observing an adhering, yet flaccid, vesicle. A more
ambitious test could be arranged by washing out the exterior solution,
replacing it by another of different ion strength but the same
osmolarity, while pinning a single vesicle for observation with a
micropipette.  Our formul\ae{} generalize readily to the case where
the ionic strength $\nhat$ is different inside and outside the
vesicle. In this way may be possible reversibly to turn adhesion
saturation on and off.

\acknowledgments
We  thank R. Bruinsma, I. Rouzina,  and B. Shklovskii for
discussions, and especially H. Aranda-Espinoza, N. Dan, T. C. Lubensky,
L. Ramos, and D. A. Weitz for an earlier collaboration leading to the
ideas presented here.
This work was supported in part by NSF grant DMR98-07156.

\section*{Appendix: Notation}\label{appa}
\subsection*{Constants}
We work in SI units. Thus the potential around a point
charge $q$ in vacuum is $\psi(r)=q/4\pi\epsilon_0 r$, where
$\epsilon_0=9\eem{12}\,$Farad/meter. We treat water as a continuum
dielectric with $\epsilon=80\epsilon_0$; inside the membrane
$\epsilon_m\approx 2\epsilon_0$.
The Bjerrum length in water is $\lB=e^2/4\pi\epsilon\kbt$; thus
$4\pi\lB=8.7\,$nm.

\subsection*{Parameters}
We take for illustration a typical ambient salt
concentration of $\nhat=1\,$mM=$6\eem4\,$nm$^{-3}$.
Then the inverse Debye length is
$\kappa=\sqrt{2\hat ne^2/\epsilon\kbt}=\sqrt{\nhat/{\rm mM}}/(9.8\,$nm). The
salt
concentration inside the vesicle is the same, due to osmotic clamping.

We suppose a mixture of surfactants, which for simplicity
have equal area per headgroup
$a_0=0.5\,$nm$^2$. Then $\smax=2e/a_0$ is the maximum bilayer charge
density and the parameter
$\beta=2\hat na_0/\kappa=0.006$ measures the relative importance of
mixing-entropy and electrostatic effects.

We use a typical artificial bilayer capacitance of
$c=0.01\,$pF/$\mu$m$^2$, 
which enters only in combination with the membrane thickness
$t$ \via{} $\tau=t\kappa\epsilon/\epsilon_{\rm m}\approx
7\sqrt{\nhat/{\rm mM}}$.

For illustration we take the experimentally-controllable mole fraction
of charged surfactants to be
$\spavb=1/2$ and one half of the corresponding counterions to be
trapped on the vesicle interior, so that $\stavb=\spavb/2=1/4$. We
also take the approaching charged dielectric objects to have charge
density 50\% greater than the membrane, or  $\smb=-3/2$.

\subsection*{Variables}
We generally denote nondimensionalized quantities with a bar or tilde:
thus $\sbar\equiv\sigma/\smax$, while $\sbb=\sigma\kappa/2\hat
ne=2\sbar/\beta$. Also the free energy density $f$ gives rise to
$\fb=\kappa f/\nhat\kbt= f/(6\ee3\kbt/\mu{\rm
m}^2)\sqrt{\nhat/{\rm mM}}$, while the electrostatic potential $\psi$
gives $\psb=e\psi/\kbt$. Various contributions to $f$ include the mixing
entropy of membrane surfactants $\fm$ and the correlation contribution
$f_c$.

The charge density $\sigma$ of a surface determines its Gouy-Chapman length
$\lambda_{\rm GC}=2\epsilon\kbt/e\sigma$. Various charge densities in
the text are defined in \tfig{fsetup}, for example
$\sigma_t=\sigma\one+\sigma_+$. $m$ denotes the 2d {\it number} density
of ions in the dense correlated cloud near a surface. $m$ in turn
determines the plasma parameter $\Gamma\equiv\lB\sqrt{\pi m}$.

Geometrical quantities include the gap width $\ell$, the total
membrane area $A$, the fraction $\gamma$ of $A$ in the adhesion zone
and its equilibrium value $\gamma_*$. The distance $z$ from a surface
is sometimes expressed using
$\zeta=\ex{\kappa z}$.

\yesPR{\bibliographystyle{prsty}}
\notPR\bibliographystyle{unsrt}


\yesPR{

\bigskip
\begin{figure}\caption{\protect\capone}\label{fsetup}\end{figure}
\begin{figure}\caption{\protect\captwo}\label{ffthick}\end{figure}
\begin{figure}\caption{\protect\capthree}\label{fgapgeom}\end{figure}
\begin{figure}\caption{\protect\capfour}\label{ffselfPBDH}\end{figure}
\begin{figure}\caption{\protect\capfive}\label{ffbartim3D}\end{figure}
\begin{figure}\caption{\protect\capsix}\label{fthreecurves}\end{figure}
\begin{figure}\caption{\protect\capseven}\label{fthickmemb}\end{figure}
\begin{figure}\caption{\protect\capeight}\label{ftotsuji}\end{figure}

\newpage
\begin{center}{\epsfysize=4truein \epsfbox{f1setup.eps}}
\end{center}\bigskip\noindent Figure 1

\newpage
\begin{center}{\epsfysize=2truein \epsfbox{f2fthick.eps}}
\end{center}\bigskip\noindent Figure 2

\newpage
\begin{center}{\epsfysize=2truein \epsfbox{f3gapgeom.eps}}
\end{center}\bigskip\noindent Figure 3

\newpage
\begin{center}{\epsfysize=2truein \epsfbox{f4fselfPBDH.eps}}
\end{center}\bigskip\noindent Figure 4

\newpage
\begin{center}{\epsfysize=3truein \epsfbox{f5fbartim3D.eps}}
\end{center}\bigskip\noindent Figure 5

\newpage
\begin{center}{\epsfysize=2.7truein \epsfbox{f6threecurves.eps}}
\end{center}\bigskip\noindent Figure 6

\newpage
\begin{center}{\epsfysize=2.5truein \epsfbox{f7thickmemb.eps}}
\end{center}\bigskip\noindent Figure 7

\newpage
\begin{center}{\epsfysize=2truein \epsfbox{f8totsuji.eps}}
\end{center}\bigskip\noindent Figure 8

} 

\end{document}